\documentclass[prd,nofootinbib,12pt]{revtex4}

\usepackage{lineno,hyperref}
\usepackage{amsfonts,amssymb,amsmath}
\usepackage{epsfig}
\usepackage{mathrsfs}
\usepackage{amssymb}
\usepackage{graphicx}
\usepackage{amsmath}
\usepackage{graphicx}

\begin{document}

\title{Revised $f_{\rm NL}$ parameter in a curvaton scenario}
\author{Lei-Hua Liu$^{1}$}
\author{Bin Liang$^{2}$}
\author{Ya-Chen Zhou$^{1}$}
\author{Xiao-Dan Liu$^{1}$}
\email{liuleihua8899@hotmail.com}
\author{Wu-Long Xu$^{3}$}
\author{Ai-Chen Li$^{3,4}$}
\affiliation{1. Department of Physics, College of Physics,
	Mechanical and Electrical Engineering,
	Jishou University, Jishou 416000, China
\\2. 	College of Civil Engineering, Hunan University of Technology, Zhuzhou, 412007, China\\
3. Institute of Theoretical Physics, Beijing University of Technology, Beijing 100124, China\\
4. Departamento de matematica da Universidade de Aveiro and CIDMA,
Campus de Santiago, 3810-183 Aveiro, Portugal
}

\begin{abstract}
We revise the Non-Gaussianity of canonical curvaton scenario with a generalized $\delta N$ formalism, in which it could handle the generic potentials. In various curvaton models, the energy density is dominant in different period including the secondary inflation of curvaton, matter domination and radiation domination. Our method could unify to deal with these periods since the non-linearity parameter $f_{\rm NL}$ associated with Non-Gaussianity is a function of equation of state $w$. We firstly investigate the most simple curvaton scenario, namely the chaotic curvaton with quadratic potential. Our study shows that most parameter space satisfies with observational constraints. And our formula will nicely recover the well-known value of $f_{\rm NL}$ in the absence of non-linear evolution. From the micro origin of curvaton, we also investigate the Pseudo-Nambu-Goldstone curvaton. Our result clearly indicates that the second short inflationary process for Pseudo-Nambu-Goldstone curvaton is ruled out in light of observations. Finally, our method sheds a new way for investigating the Non-Gaussianity of curvaton mechanism, espeically for exploring the Non-Gaussianity in MSSM curvaton model. 

\end{abstract}

\maketitle

\section{Introduction}

In traditional diagram of producing the curvature perturbation, it is sourced by the quantum fluctuations of inflationary field. In these broad class of single inflationary field theories, it experiences some initial condition problems associated with its corresponding potential. In order to relax these restrictions of single field inflation, one nice alternative called curvaton mechanism was proposed \cite{Enqvist:2001zp, Lyth:2001nq, Moroi:2001ct}, in which the energy density of curvaton is subdominant comparing with inflaton's during inflationary period. After inflation decay, the role of curvaton will be more and more significant producing the isocurvature perturbation, which can be transferred into curvature perturbation seeding the temperature fluctuation on cosmological microwave background (CMB).

Due to the appearance of CMB, there are huge data waiting for the investigations. In particular, the most common method is calculating the power spectrum of scalar field (driving the curvature perturbation) characterizing by two point correlation function, its corresponding spectral index and tensor to scalar ratio. However, most data are still mysterious expecting a new theoretical method for exploring these treasures. Under this background, the calculation of Non-Gaussianity (NG) identified with three point correlation function was proposed \cite{Maldacena:2002vr}. Combining with curvaton scenario, NG, associated with its fraction of energy density among total energy density, could also be produced as curvaton dominates over the energy density \cite{Bartolo:2003jx,Lyth:2005fi,Bartolo:2004ty}. Upon relaxing this condition (curvaton dominates over energy density), it could yield large NG \cite{Lyth:2002my}. However, current observation constrains these models \cite{Ade:2015ava}, namely characterizing by the local non-linearity parameter $f_{\rm NL}$ that cannot be large. This local $f_{\rm NL}$ is suppressed by the quadratic potential plus quartic potential \cite{Mukaida:2014wma} and also in string axionic potential \cite{Dimopoulos:2011gb,Kawasaki:2012gg}. Furthermore, the observable $f_{\rm NL}$ also puts an enhanced constraint for the decay epoch of curvaton and its field value at the horizon exit \cite{Sharma:2019qan}. The implications of NG features in curvaton scenario were also studied in Refs. \cite{Huang:2008ze,Huang:2008bg,Gong:2009dh}. On contrary, NG could be produced in various curvaton models \cite{Harigaya:2012up,Enomoto:2012uy,Fonseca:2012cj}.

In most curvaton scenarios, curvaton usually is considered as an independent field. If taking the thermal effects into account, the large NG is the necessary product due to the observed curvature perturbation \cite{Mukaida:2014yia}, even the curvaton could be realized in low energy inflation comparing with tradiational curvaton mechanism \cite{Dimopoulos:2004yb}. Furtherly, a samilar curvaton mechanism could also be achieved due to the coupling between the inflaton and curvaton \cite{Rodriguez:2004yc}. From another perspective of independent curvaton, it naturally embeds into two field inflationary theory, in which it could produce the sizable NG within observations \cite{Enomoto:2013bga}. Very recently, Ref. \cite{Liu:2020zzv} rigorously realizes the curvaton mechanism under the covariant framework of field space. Taking the curvaton and inflaton into account for the perturbation, NG could be generated by inflaton curvaton mixed model \cite{Fonseca:2012cj}, even the curvaton could drive the second inflationary process \cite{Byrnes:2014xua}. However, current observational constraints are not capable for distinguishing between the inflaton-curvaton mixed model and single field inflation \cite{Kinney:2012ik}. As curvaton explicitly couples to the super-heavy matter, it will lead to observational signal including NG \cite{Li:2020xwr,Kumar:2019ebj}. From another aspect, curvaton is dubbed as some scalar fields, {\it i.e.}, Pseudo-Nambu-Goldstone Boson or right-handed sneutrino curvaton {etc} \cite{Haba:2017fbi, Dimopoulos:2003az, BasteroGil:2002xr, Enqvist:2002rf,Lee:2011dy}. Due to the unification of string theory, the curvaton scenario could also be applied into the string cosmology framework \cite{Li:2008fma,Zhang:2009gw}, in which it yields considerable NG. Since the energy scale of inflation is far from Planck scale. Curvaton scenario could be embedded into  the minimal supersymmetric Standard Model \cite{Mazumdar:2011xe}. Another origin comes via the inflaton decay \cite{Byrnes:2016xlk}.

The NG is associated with three point correlation function, in order to investigate the NG, $\delta N$ formalism was proposed \cite{Lyth:2005fi} depending on the surface of energy density slicing. Its huge merit is only needing the relation of the corresponding background field and e-folding number. Based on previous work, $\delta N$ formalism was systematically developed by \cite{Sasaki:2006kq}. $\delta N$ formalism has become a standard procedure to evaluate the power spectrum and NG in the multi-field inflationary framework including the curvaton scenario (the canonical kinetic term of field space). Ref. \cite{Cai:2010rt} modified the $\delta N$ formalism at the slice of curvaton energy density, their method could proceed the curvaton mechanism in various periods (matter domination, radiation domination, second inflationary period) explicitly associated with equation of state $w$ (EoS). However, this traditional $\delta N$ formalism cannot analytically evaluate the various curvaton models (distinct potentials). In order to compensate this flaw, Refs \cite{Kawasaki:2011pd,Kobayashi:2012ba} also proposed a modified $\delta N$ formalism, in which this method could deal with various curvaton potentials analytically in principle. However, they assumed that different periods have simple attractor solution characterizing by an ordinary parameter $c$, in which the kinetic term is neglected and its contribution is enrolled into this parameter. This estimation of their method is too coarse comparing to traditional calculation. The best way is including the contribution EoS $w$ since it is model independent. In light of above theoretical motivations, we suggest a generalized $\delta N$ formalism unified to evaluate the Non linearity parameter $f_{\rm NL}$.

This paper is organized as follows. In section \ref{the generalized delta n}, we will revise the $\delta N$ formalism based on \cite{Sasaki:2006kq} and meanwhile we also give our central formula of Non linearity parameter $f_{\rm NL}$. In section \ref{case study}, we study the most classical curvaton model whose potential is quadratic and Pseudo-Nambu-Goldstone curvaton. Section \ref{conclusion} gives our main conclusions.

 All of the calculations are adopted in the natural units which $G=M_P=c=1$, where $G$ is the Newton constant, $M_P$ is the Planck mass and $c$ is the speed of light.


\section{The generalized $\delta N$ formalism of curvaton decay}
\label{the generalized delta n}
In this section, we will generalize the $\delta N$ formalism. In light of Ref. \cite{Cai:2010rt} and Ref.~\cite{Kawasaki:2011pd,Kobayashi:2012ba}, our extending framework contains their merits. The main advantage of \cite{Kawasaki:2011pd,Kobayashi:2012ba} for $f_{NL}$ is that they build the explicit relation of the onset of oscillation of curvaton and the curvaton value as inflation ends, which is not included in a traditional $\delta N$ formalism. As for Ref. \cite{Cai:2010rt}, they construct $f_{NL}$ associated with equation of state $w$ (EoS) except the fraction of curvaton energy density to total energy density denoted by $r_{\rm decay}$. Firstly, we will review the $\delta N$ formalism.

\subsection{Recap of $\delta N$ formalism for curvaton decay}
\label{Recap of delta n}
In a traditional curvaton scenario, it will generate the Non-Gaussianity essentially characterizing by non-local Nonp-Gaussianity parameter $f_{\rm NL}$. In order to obtain its explicit formula, the most common method for copying with is so-called $\delta N$ formalism \cite{Sasaki:2006kq} since it only requires the relation between the background field and e-folding number.The curvature perturbation can be expanded as order by order,
\begin{equation}
\zeta(x)=\zeta_1(x)+\frac{1}{2!}\zeta_2(x)+\frac{1}{3!}\zeta_3(x),
\label{cur perturbation}
\end{equation}
with $\zeta_2=\frac{6}{5}f_{\rm NL}\zeta_1^2$ and $\zeta_3=\frac{54}{25}g_{\rm NL}\zeta_1^3$, where $\zeta_1$ is explicitly proportional to Gaussian field, $\zeta_2$ and $\zeta_3$ are related to Non-Gaussian field associated with Gaussian field for non-Local Non-Gaussianity parameter $f_{\rm NL}$ and $g_{\rm NL}$. Here, we only concern with $f_{\rm NL}$ since $g_{\rm NL}\propto f_{\rm NL}^2$ and it will be suppressed at higher order. $f_{\rm NL}$ originates from the three point correlation functions,
\begin{equation}
\langle \zeta(k_1)\zeta(k_2)\zeta(k_3)\rangle =(2\pi)^3B(k_1,k_2,k_3)\delta^3(\sum_{n=1}^3k_n),
\label{three point}
\end{equation}
where $B(k_1,k_2,k_3)=\frac{5}{6}f_{\rm NL}(P(k_1)P(k_2)+2~\mathbf{perm}.)$ with $P(k_i)$ is the power spectrum of $\zeta_{k_i}$ field.

In order to relate to e-folding number $N$, once adopting uniform density hypersurfaces of curvaton, then the curvature perturbation can be denoted in terms of non-linear curvature perturbation,
\begin{equation}
\zeta(x)= \delta N(x)+\frac{1}{3}\int_{\bar{\rho}(t_0)}^{\rho(x)}\frac{d\tilde{\rho}}{\tilde{\rho}+\tilde{P}},
\label{nonlinear perturbation}
\end{equation}
where $\delta N$ is the perturbed expansion, $\tilde{\rho}$ is the local energy density and $\tilde{P}$ is the local pressure. Given that the curvaton decay occurs in matter domination period (MD), then one naturally neglects the contribution of pressure. Subsequently, integrating both sides of Eq. (\ref{nonlinear perturbation}) and choosing flat slice, one obtains
\begin{equation}
\rho_\chi=\bar{\rho}_\chi\exp(3\zeta_\chi).
\label{integration of cur}
\end{equation}
For curvaton field, its perturbation can be defined by,
\begin{equation}
\chi_*=\bar{\chi}+\delta_1\chi_*,
\label{perturbation of curvaton field}
\end{equation}
where $\delta_1\chi_*$ denotes the vacuum fluctuations of curvaton field. For depicting the curvature perturbation of curvaton, we need relating the Hubble crossing value to the initial amplitude of curvaton oscillation. In order to achieve this goal, one could use the Taylor expansion to build their relation,
\begin{equation}
g(\chi_*)=g(\bar{\chi}+\delta_1\chi_*)=\bar{g}+\sum_{n=1}^\infty\frac{g^{(n)}}{n!}\bigg(\frac{\delta_1\chi_{\rm osc}}{g'}\bigg)^n,
\label{relation of cur}
\end{equation}
where $g'=\frac{dg}{d\chi_*}$ and $\chi_{\rm osc}$ denotes the value of curvaton field that begins to oscillate. Apparently, $g(\chi_*)$ depends on the model. Until present, the discussion for curvature perturbation of curvaton is generic which means that the curvaton potential is general. In order to relate to some specific curvaton models, Ref. \cite{Sasaki:2006kq} assumes that the simplest potential (quadratic potential) for curvaton. Apparently, it shows that
$g(\chi_*)\propto \chi_*$. Subsequently, one can consider this potential as energy density and then expand them to the second order of perturbation of curvaton field for comparing, finally we find that
\begin{eqnarray}
\zeta_{\chi 1}=\frac{2}{3}\frac{\delta_1\chi}{\bar{\chi}},\\
\zeta_{\chi 2}=-\frac{3}{2}\big(1-\frac{g g"}{g'^2}\big).
\label{relation of cur perturbation}
\end{eqnarray}
Nextly, we need to find the relation between $\zeta_\chi$ and $\zeta$. Following the sudden decay approximation, this relation can be analytically obtained, which is realized on a uniform total density hypersurface as $H=\Gamma_\chi$ (the decay rate of curvaton). On this curvaton decay hypersurface, one accordingly have
\begin{equation}
\rho_r(t_{\rm decay})+\rho_\chi(t_{\rm decay})=\bar{\rho}(t_{\rm decay}),
\label{relation of energy density on total}
\end{equation}
where $\bar{\rho}$ denotes the background field energy density. Meanwhile, we have $\delta N=\zeta$ on the curvaton decay hypersurface. Observing that the production of curvaton decay is relativistic and total pressure $P=\frac{1}{3}\rho$, consequently one easily obtains that
\begin{eqnarray}
\rho_r=\bar{\rho}_r\exp[4(\zeta_r-\zeta)],\\
\rho_\chi=\bar{\rho}_\chi\exp[3(\zeta_\chi-\zeta)].
\label{formula of rho}
\end{eqnarray}
Using these two formulas into Eq. (\ref{relation of energy density on total}) and defining a dimensionless quantity
$\Omega_\chi=\bar{\rho}_\chi/(\bar{\rho}_\chi+\bar{\rho}_r)$, after some algebra, one obtains that
\begin{equation}
(1-\Omega_\chi)\exp[4(\zeta_r-\zeta)]+\Omega_\chi\exp[3(\zeta_\chi-\zeta)]=1.
\label{central formula}
\end{equation}
Once deriving this central formula of $\delta N$ formalism, we can set the relations between the $\zeta_\chi$ and $\zeta$. Expanding up to the second order of Eq. (\ref{central formula}), we collect these relations,
\begin{eqnarray}
\zeta_1=r_{\rm decay}\zeta_{\chi1},\\
\zeta_2=\bigg[\frac{3}{2r_{\rm decay}}\big(1+\frac{gg''}{g'^2}\big)-2-r_{\rm decay}\bigg]\zeta_{\chi2}^2,
\label{relation of zeta}
\end{eqnarray}
where we have defined
\begin{equation}
r_{\rm decay}=\frac{3\Omega_{\chi,\rm decay}}{4-\Omega_{\chi,\rm decay}}=\frac{3\bar{\rho}_\chi}{3\bar{\rho}_\chi+4\bar{\rho}_r}.
\label{rdecay}
\end{equation}
It naturally yields non-linearity parameter using the sudden decay approximation \cite{Lyth:2005fi,Bartolo:2004ty},
\begin{equation}
f_{\rm NL}=\frac{5}{4r_{\rm decay}}\big(1+\frac{gg''}{g'^2}\big)-\frac{5}{3}-\frac{5r_{\rm decay}}{6}.
\label{fnl}
\end{equation}
Observing that this non-linearity parameter highly depends on the $r_{\rm decay}$, meanwhile mildly depends on the structure of model showing in $g$ and $g'$. Although we adopted the simplest potential for curvaton, the final result is almost quadratic potential independent. Actually, one can roughly estimate this result since when expanding the energy density of curvaton up to the second order. Subsequently, one can discover via Eq. (\ref{integration of cur},~\ref{relation of cur}) that the background of curvaton will be cancelled as comparing them through their equation.

Furthermore, the generic potential of curvaton should be taken into account. The time of occurrence of curvaton mechanism (various decays of curvaton models will happen in RD or MD) is also different. In order to compensate these two missing places into curvaton mechanism, some distinct generalized $\delta N$ formalisms are proposed.

\subsection{Generalized $\delta N$ formalism}
\label{generalized delta n}
In this section, we will construct a generalized $\delta N$ formalism with a generic potential and EoS $w$. Consequently, it is valid for broad kinds of curvaton models. In Ref. \cite{Cai:2010rt}, they innovatively assumed that the curvaton decay occurs on a uniform curvaton density slice. Being different with definition of total energy density in section \ref{Recap of delta n}, they found that
\begin{equation}
\zeta=\zeta_\chi+\frac{1}{4}\ln\big(\frac{4\bar{\rho}_r+3(\bar{\rho}_\chi+\bar{P}_\chi)}{4\rho_r+3(\bar{\rho}_\chi+\bar{P}_\chi)}\big).
\label{zeta chi}
\end{equation}
By inserting Eq. (\ref{formula of rho}) into Eq. (\ref{zeta chi}), they obtained
\begin{equation}
\bigg(1-\frac{1-3w}{4}\Omega_\chi\bigg)\exp[4(\chi-\chi_r)]=(1-\Omega_\chi)\exp[4(\zeta_r-\zeta_\chi)]+\frac{3(1+w)}{4}\Omega_\chi,
\label{central formula of chi}
\end{equation}
where they defined $w=\frac{\bar{P}_\chi}{\bar{\rho}_\chi}$. Following the standard procedure, the relation between the $\zeta$ and $\zeta_\chi$ can be derived order by order,
\begin{eqnarray}
\zeta_1=\tilde{r}_{\rm decay}\zeta_{\chi1},\\
\frac{\zeta_2}{\zeta_{\chi2}^2}=\frac{3(1+w)}{2\tilde{r}_{\rm decay}}\bigg(1+\frac{gg''}{g'^2}\bigg)+\frac{1-3w}{\tilde{r}_{\rm decay}}-4,
\label{relation of zeta and zetachi in chi}
\end{eqnarray}
where $\tilde{r}_{\rm decay}=\frac{3(1+w)\Omega_\chi}{4+(3w-1)\Omega_\chi}$ is introduced. Apparently, the non-linearity parameter associated with non-Gaussianity can be explicitly derived by,
\begin{eqnarray}
f_{\rm NL}=\frac{5}{4}\frac{1+w}{\tilde{r}_{\rm decay}}\bigg(1+\frac{gg''}{g'^2}\bigg)+\frac{5}{6}\frac{1-3w}{\tilde{r}_{\rm decay}}-\frac{10}{3}.
\label{fnl for cur density}
\end{eqnarray}
Observing that the value of $f_{NL}$ will be enhanced in the limit of $w\rightarrow 0$ which is equivalent to $\tilde{r}_{\rm decay}\rightarrow 0$. Ref. \cite{Cai:2010rt} has noticed that this case will be appeared in the secondary inflation. The similar process was also discussed in various curvaton models \cite{Cai:2009hw,Huang:2008zj}. Consequently, one can conclude that $w$ is a possible criteria for assessing the occurrence of secondary inflationary process.

This non-linearity parameter is tiny different comparing to (\ref{fnl}). This difference comes from the slice of energy density. In inflationary period, there are at least two components if requiring the existence of curvaton field. In order to remove the influence of other field to the non-Gaussianity, this method is necessary and more precise comparing to traditional $\delta N$ formalism. The huge merit of this generalized $\delta N$ formalism is that it could deal with the second scalar field in different epoches. Especially for the curvaton mechanism, it is usally dubbed as pressless matter, namely that it happens in MD before it decays. Within the introduction of this method, the curvaton mechanism could be fullfilled in various era. Consequently, it will extend the application of curvaton mechanism.

However, one cannot manage it analytically with generic potential besides the quadratic potential. Ref. \cite{Kobayashi:2012ba,Kawasaki:2011pd} accordingly proposed another generalized $\delta N$ formalism for dealing with the generic potential analytically. In their method, the non-linearity parameter is written by
\begin{align}
f_{\rm NL}
&=  -\frac56 r_{\rm decay} - \frac53 + \frac{5}{2r_{\rm decay}} (1 +   A),
\label{fNL tomo}
\end{align}
where $A$ is given by
\begin{align}
A &=
 \left[   \frac{V'(\chi_{\rm osc})}{V (\chi_{\rm osc}) } -  \frac{3 X(\chi_{\rm osc})}{\chi_{\rm osc}} \right]^{-1}
 \left[\frac{X'(\chi_{\rm osc})}{1 - X(\chi_{\rm osc}) }
  +  \frac{V^{''}(\chi_{\rm osc})}{V' (\chi_{\rm osc}) } - \left( 1 - X(\chi_{\rm osc}) \right) \frac{V^{''}(\chi_\ast)}{V' (\chi_{\rm osc}) } \right]
 \notag \\
 &
  +\left[  \frac{V'(\chi_{\rm osc})}{V (\chi_{\rm osc}) } -  \frac{3 X(\chi_{\rm osc})}{\chi_{\rm osc}} \right]^{-2}
\left[
 \frac{V^{''}(\chi_{\rm osc})}{V (\chi_{\rm osc}) } - \left( \frac{V'(\chi_{\rm osc})}{V (\chi_{\rm osc}) } \right)^2  -  \frac{3 X'(\chi_{\rm osc})}{\chi_{\rm osc}}
 +  \frac{3 X(\chi_{\rm osc})}{\chi_{\rm osc}^2}  \right] \, .
\label{A}
\end{align}
Here $A$ is characterized by a curvaton with a generic energy potential, in which it experiences a non-uniform onset of its oscillation. Its validity only requires starting a sinusoidal oscillation as satisfying with
\begin{equation}
H_{\rm osc}^2 = \frac{V'(\chi_{\rm osc})}{c \chi_{\rm osc}}
\label{onset_osc2}
\end{equation}
where $c$ is  given by $ 9/2$ and $5$ when the curvaton begins to oscillate during matter domination (MD) and radiation domination (RD), respectively. The information of different period is explicitly included in parameter $c$ characterizing by the attractor solution.

In order to relate the method of Ref. \cite{Cai:2010rt}, we need to find their correspondence between Eq. (\ref{fNL tomo}) and Eq. (\ref{fnl for cur density}). Before finding the correspondence, the relation between Eq. (\ref{fNL tomo}) and Eq. (\ref{fnl}) is necessary since these two methods are adopted in the total energy density slice. Maybe this slice for \cite{Kobayashi:2012ba,Kawasaki:2011pd} is not explicit. However, one can easily check that the whole calculation is depending on the total energy density in the curvaton dominant period after inflation. Furthermore, the total energy slice is approximately equalled to the curvaton energy density slice after inflation, since the curvaton is dominant which is also an assumption for original curvaton scenario. In light of this logic, we should find the correspondence between Eq. (\ref{fNL tomo}) and Eq. (\ref{fnl}) and then explicitly adopt this correspondence for Eq. (\ref{fNL tomo}). Comparing with Eq. (\ref{fNL tomo}) and Eq. (\ref{fnl}), an explicit correspondence can be found by
\begin{equation}
1+2A=\frac{gg''}{g^{'2}}.
\label{correspondence of A}
\end{equation}
Using this correspondence into Eq. (\ref{fnl for cur density}), we obtain
\begin{eqnarray}
f_{\rm NL}=\frac{5}{2}\frac{1+w}{\tilde{r}_{\rm decay}}\bigg(1+A\bigg)+\frac{5}{6}\frac{1-3w}{\tilde{r}_{\rm decay}}-\frac{10}{3}.
\label{fnl for cur density1}
\end{eqnarray}
In this formula, we observe that $\tilde{r}_{\rm decay}$ is also the function of $w$. Following the traditional logic, we will work with $f_{\rm NL}$ in terms of $r_{\rm decay}$ and $w$. In order to achieve this goal, the relation between $r_{\rm decay}$ and $\tilde{r}_{\rm decay}$ is mandatory. In light of their relation, the non-linearity parameter can be rewritten by
\begin{eqnarray}
f_{\rm NL}=\frac{5 (3 A w+3 A+4)}{6 r_{\rm decay} (w+1)}+\frac{5 \left(3 A w^2+3 A w-4\right)}{6 (w+1)}.
\label{fnl for cur density2}
\end{eqnarray}

Thus, we obtain the central result of this paper, in which it could tackle the generic potential analytically and it could assess the existence of second inflationary process for curvaton field. In the next section, we will investigate the non-linearity parameter $f_{\rm NL}$ in various curvaton models under the observational constraints.

\section{Case study}
\label{case study}
The realization of curvaton mechanism depends on the models, particularly it depends on the potential of curvaton. The shape of potential for curvaton will lead to the difference in various curvaton models, $\rm i.g.$ chaotic curvaton model, axionic curvaton, $\rm e.t.c.$

Before discussing the Non-Gaussianity identified with non-linearity parameter $f_{\rm NL}$.  The consideration of power spectrum of curvaton must be taken into account. Recalling that our derivation of $f_{\rm NL}$ is mainly according to the framework of \cite{Kawasaki:2011pd,Kobayashi:2012ba}, they found that the power spectrum of curvaton is nearly scale invariant in different values of $k$ for various models of curvaton (exactly speaking for the various potentials of curvaton). Furthermore, Ref. \cite{Liu:2019xhn} also studied that power spectrum is only depending on $r_{\rm decay}$ and $\chi$ explicitly. Thus, the power spectrum of curvaton is the same for various models of curvaton. This issue can be easily checked in \cite{Sasaki:2006kq,Kawasaki:2011pd,Kobayashi:2012ba}.

The second issue should clarified, which is related to the period of occurrence for curvaton mechanism. In traditional curvaton mechanism, it happens in the MD whose corresponding value of $w=0$ behaving like a presssureless matter \cite{Enqvist:2001zp, Lyth:2001nq, Moroi:2001ct}. However, this similar mechanism can be realized in different periods,  $\it i.g.$, Ref. \cite{Liu:2019xhn} has considered a curvaton mechanism occurred in RD due to the decay of inflaton inspired by \cite{Jiang:2018uce,Cai:2011zx}, in which the key ingredient is the explicit coupling between the curvaton field and inflaton field. Once obtaining the curvaton, the curvaton field will also decay into Standard Model degrees of freedom as inflato decay (the generation of curvaton comes via inflaton decay). Consequently, it will lead to amount of isocurvature perturbation without thermalising with the Standard Model degrees. From the current cosntraint \cite{Ade:2015ava}, the power spectrum of isocurvature perturbation comparing with curvature perturbation cannot be large. In order to tranfer this isocurvature perturbation into curvature perturbation, Refs. \cite{Allahverdi:2006dr,Enqvist:2002rf} proposed the viable curvaton mechanism embedded into MSSM in light of \cite{Dvali:2003em} by considering the thermalization.  Thus, the curvan mechanism can be realized in various epoches under the framework of MSSM. Furthermore, curvaton mechanism can also be achieved by the curvaton brane leading to the large NG \cite{Cai:2010rt} whose corresponding value of $w=-1$. From the central formula of (\ref{fnl for cur density}), it is also known that our proceeding with the curvaton mechanism is adopted for various epoches. 
Meanwhile, the variants of $\delta N$ formalism contains the method proceeding with the curvaton mechanism in distinct periods. In light of the above theoretical motivations, we could find that the curvaton mechanism will be realized in various periods corresponding to different values of $w$.

\subsection{Chaotic curvaton}
Chaotic curvaton indicates that the potential of curvaton is quadratic. These kinds of curvaton have been investigated broadly, in particular, for the non-Gaussianity characterizing by non-linearity parameter $f_{\rm NL}$ \cite{Lyth:2005fi, Bartolo:2004ty}. In light of quadratic potential, Ref. \cite{Sasaki:2006kq} proposed a generalized $\delta N$ formalism to investigate the non-Gaussianity, in which curvature perturbation can be derived up to any order. We accordingly concern the second order of curvature perturbation associated with $f_{\rm NL}$.

We will give a analysis of $f_{\rm NL}$ for chaotic curvaton based on our central result (\ref{fnl for cur density1}). In our previous work \cite{Liu:2019xhn}, we clearly show that $A=-\frac{1}{2}$ as the potential of curvaton proportional to $\chi^2$ where $\chi$ denotes the value of curvaton field, in which it is explicitly consistent with simple analysis of Ref. \cite{Kobayashi:2012ba} (only adopting the different notation for the fraction of curvation energy density among the total energy density). Accordingly, the central result for $f_{\rm NL}$ becomes
\begin{equation}
f_{\rm NL}=-\frac{5 (3 w-5)}{12 r_{\rm decay} (w+1)}-\frac{5 \left(3 w^2+3 w+8\right)}{12 (w+1)}.
\label{cenral fnl with quadratic potential}
\end{equation}
We will use this formula for investigating the non-Gaussianity comparing to previous relevant work. This $f_{\rm NL}$ is a generic formula for curvaton associated with Non-Gaussianity.

Case $\mathbf{a}$: $\rm \mathbf{w\rightarrow -1}$

In various models, the EoS $w$ could have different values. In Ref. \cite{Cai:2010rt}, they constructed a curvaton scenario under the framework of brane world, in which the corresponding $w\rightarrow -1$. In this case, it clearly indicates that $f_{\rm NL}$ will be divergent exceeding the range of current observational constraints \cite{Akrami:2018odb}.

Case $\mathbf{b}$: $\rm \mathbf{w\rightarrow 0}$

In this case, the curvaton behaves as the pressureless matter. $f_{\rm NL}$ simplifies into
\begin{equation}
f_{\rm NL}=\frac{25}{12 r_{\rm decay}}-\frac{10}{3}.
\label{cenral fnl with quadratic potential case b}
\end{equation}
In limit of $r_{\rm decay}\rightarrow 1$, $f_{\rm NL}=-\frac{5}{4}$ which nicely recovers with Eq. (26) in Ref. \cite{Sasaki:2006kq} in the absence of non-linear evolution for the curvature perturbation of curvaton (also emphasized in Ref. \cite{Lyth:2005fi}), in which the curvaton scenario is the simplest curvaton model whose potential is $\frac{1}{2}m_\chi^2\chi^2$ ($\chi$ denotes the curvaton field) and behaves as pressureless matter according to our analysis. Meanwhile, curvaton dominates the energy density. For large Non-Gaussiantiy, it requires that $r_{\rm decay}\rightarrow 0$. In order to better understand the possible range of $r_{\rm decay}$, we will plot Eq. (\ref{cenral fnl with quadratic potential case b}).

Case $\mathbf{c}$ : $\rm \mathbf{w\rightarrow \frac{1}{3}}$

In this case, curvaton decay is a relativistic process. Then, $f_{\rm NL}$ becomes
\begin{equation}
f_{\rm NL}=\frac{5}{4 r_{\rm decay}}-\frac{35}{12}.
\label{cenral fnl with quadratic potential case c}
\end{equation}
A similar analysis will be given as in case b. In limit of $r_{\rm decay}\rightarrow 1$, $f_{\rm NL}\rightarrow -\frac{5}{3}$. The value is almost the same with case b, in which one cannot distinguish the tiny difference between case $b$ and case $c$. Frankly speaking, curvaton is an independent and extra field during inflationary process (even including the preheating process), however curvaton could be induced by the inflaton decay whose realization occurs from the transferring of entropy perturbation to curvature perturbation \cite{Liu:2019xhn}, in order to realize this transferring, the curvaton can be embedded into MSSM \cite{Allahverdi:2006dr}.

\begin{figure}[h!]
 \centering
  \includegraphics[height=7.0cm, width=7.5cm]{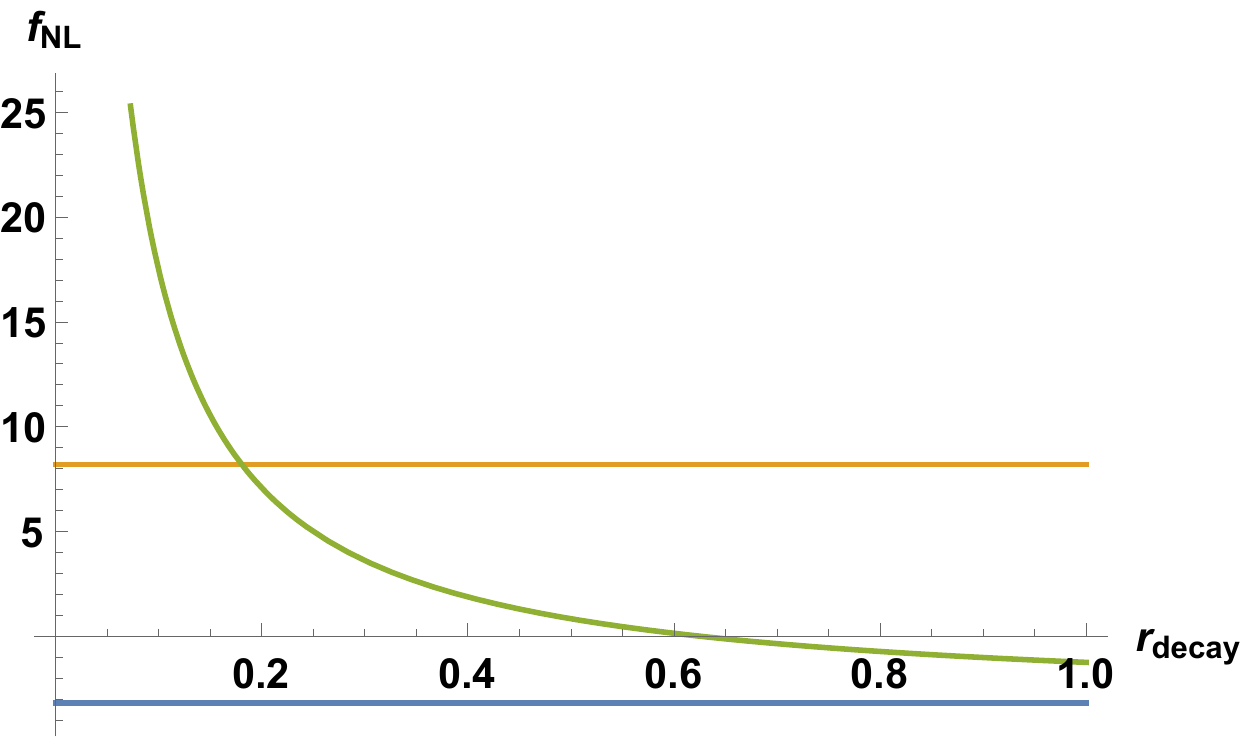}
\includegraphics[height=7.0cm, width=7.5cm ]{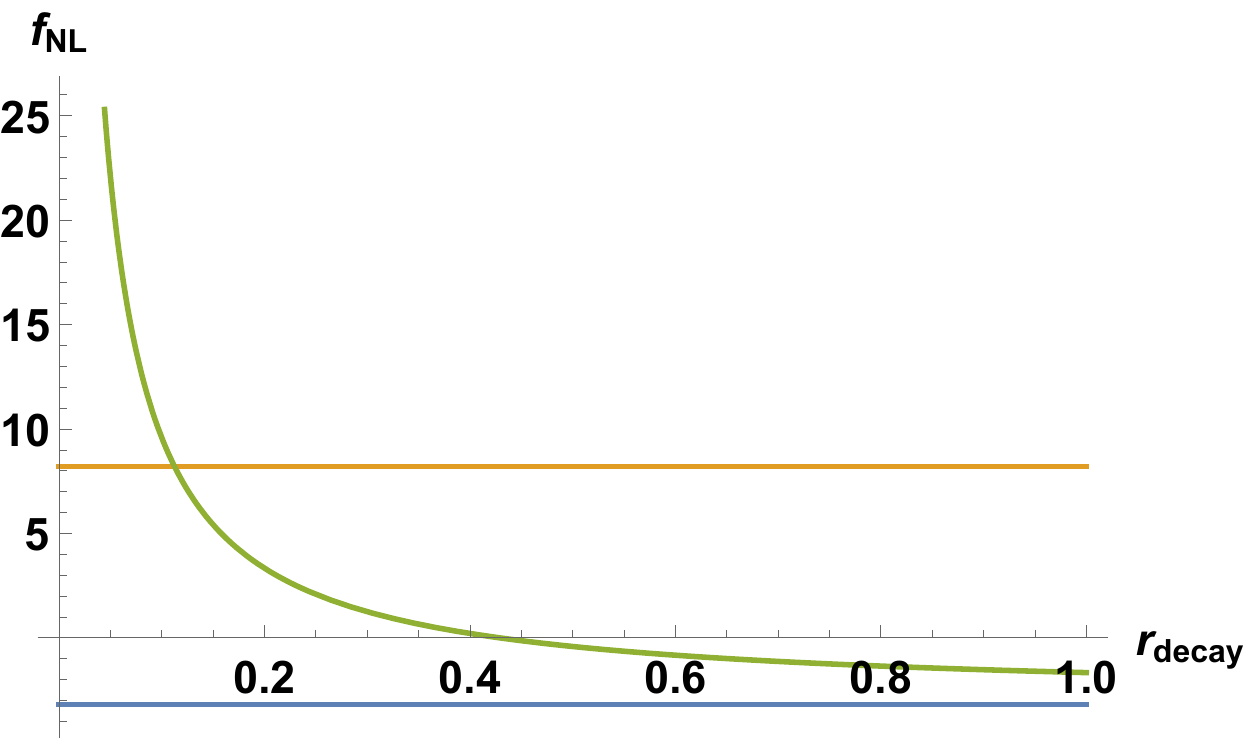}
\vskip -0.4cm
 \caption{ Left panel shows the non-linearity parameter $f_{\rm NL}$ for case $b$ and right panel illustrates the case case $c$. The brown and blue line denote the upper and lower bound for $f_{\rm NL}$. The corresponding value of $r_{\rm decay}$ is $0.18$ and $0.11$ with respect to case $b$ and case $c$, respectively. }
 \label{fnl with case b and c}
 \end{figure}

We have discussed the non-linearity parameter with various cases of chaotic curvaton, whose potential is proportional to $\chi^2$. Although we cannot distinguish the difference for case $b$ and case $c$ via observational contraints, it is expecting for obtaining the distinct values for its corresponding cases. Ref \cite{Ade:2015ava} tells that $f_{\rm NL}=2.5\pm 5.7$, afterwards, combining with Eq. (\ref{cenral fnl with quadratic potential case b},\ref{cenral fnl with quadratic potential case c}), we could plot for comparing them. In figure \ref{fnl with case b and c}, it explicitly depicts that the constraints of  $r_{\rm decay}$ for case $b$ and case $c$, respectively. The corresponding values are $0.18$ for left panel (case $b$) and $0.11$ for right panel (case $c$). This trend is logical since case $c$ illustrates curvaton behaves as relativistic matter meaning the curvaton will last longer-time occurrence of its decay.

\begin{figure}[h!]
 \centering
  \includegraphics[height=8.8cm, width=8.52cm]{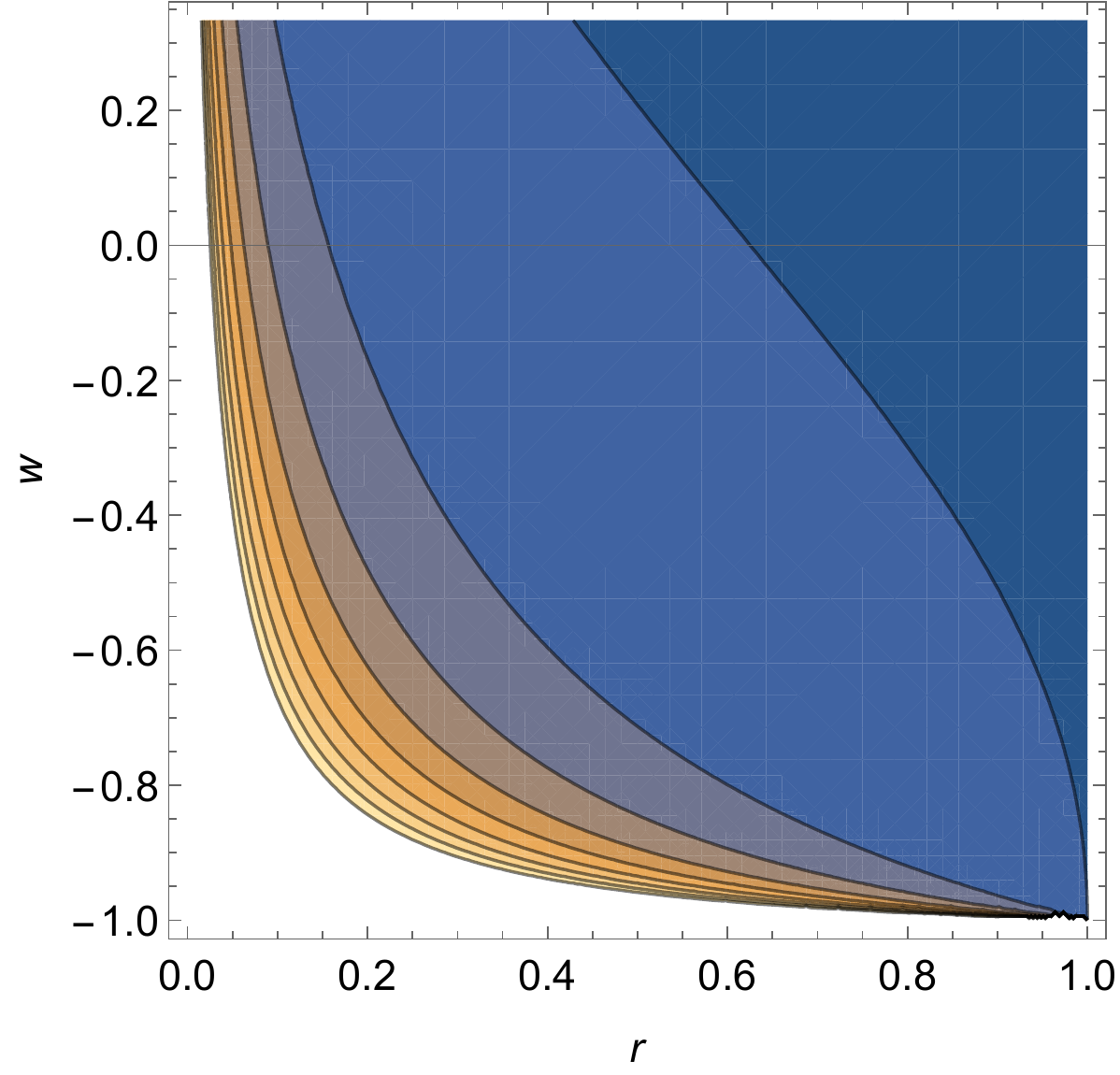}
\includegraphics[height=8.8cm, width=1.05cm ]{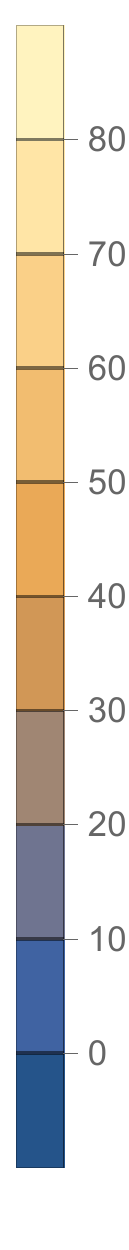}
\vskip -0.4cm
 \caption{{\it Contour plot of non-linearity parameter (\ref{cenral fnl with quadratic potential}):} The horizontal line corresponds to $r_{\rm decay}$ whose range is $0 \leqslant r_{\rm decay}\leqslant 1$ including the whole possible value. The vertical line denotes the value of equation of state $w$ locating from $-1$ to $\frac{1}{3}$, in which it includes that dark energy epoch, radiation domination period, matter domination period and it could indicate the transition from one era to another era. The right panel shows that the value of $f_{\rm NL}$ matching its corresponding color.}
 \label{fnl with quadratic potential}
 \end{figure}

For the careful reader, they may find that there is still some losing information for the transition from $w\rightarrow 0$ to $w\rightarrow \frac{1}{3}$, since the curvaton will become the relativistic matter as the longtime occurrence of curvaton decay (from MD to RD). If considering this case, $r_{\rm decay}$ will be a small number, but what the precise value is. We need the more detailed investigation of $f_{\rm NL}$ varying $w$. In order to achieve this goal, we show the density plot of non-linearity parameter $f_{\rm NL}$ depending on the parameter $r_{\rm decay}$ and $w$ in figure \ref{fnl with quadratic potential}.
It clearly indicates that the Non-Gaussianity will be dramatically enhanced as $w\rightarrow 0$ and $r_{\rm decay}\rightarrow 0$, which is consistent with our previous discussion. Interestingly, the $f_{\rm NL}$ is still within the observational constraints \cite{Akrami:2018odb}, in which $w$ is approaching $-1$ before curvaton decays. This is one of our new findings for chaotic curvaton model. As the decay of curvaton is continuing,  we find that there are lots of parameter spaces satisfied with observational constraints, showing the blue area of figure \ref{fnl with quadratic potential} as $r_{\rm decay}<0.5$.

\subsection{Pseudo-Nambu-Goldstone curvaton}

In this case, we will further consider the curvaton could origin from microscopic physics, namely, pseudo-Nambu-Goldstone boson with a broken $U(1)$ symmetry. The curvaton mass will be suppressed by the approximating symmetry. Since curvaton has the periodicity of $U(1)$ leading to minima and maxima along the potential. Therefore it will generate the blue and red tiled curvature perturbation of curvaton. What we concern is the potential of Pseudo-Nambu-Goldstone curvaton, it reads as
\begin{equation}
V(\chi)=\Lambda^4\bigg[1-\cos\left(\frac{\chi}{f}\right)\bigg],
\label{potential of axionic curvaton}
\end{equation}
where $f$ and $\Lambda$ denote the energy scale. In order to obtain its corresponding $f_{\rm NL}$, the relation between the $\chi_*$ and $\chi_{\rm osc}$ is mandatory. For achieving this goal, we need the modified
KG equation (\ref{onset_osc2}), one can derive
\begin{equation}
\ln\bigg[\frac{\tan(\chi_{\rm osc}/2f)}{\tan(\chi_{*}/2f)}\bigg]=-\frac{N_*}{3H_{\rm inf}^2}\frac{\Lambda^4}{f^2}-\frac{1}{2(c-3)}\frac{\chi_{\rm osc}/f}{\sin(\chi_{\rm osc}/f)},
\label{relation of chistar and chiosc}
\end{equation}
where $N_*$ denotes the e-folding number at the horizon exit, $H_{\rm inf}$ represents the Hubble parameter during inflation. After some algebras, we can represent $\chi_*$ in terms of $\chi_{\rm osc}$,
\begin{equation}
\chi_*=\frac{1}{f}\bigg[{\rm arccot}\bigg(\exp\bigg(-\frac{\frac{3 f \chi_{\rm osc} \csc \left(\frac{\chi_{\rm osc}}{f}\right)}{c-3}+\frac{2 \Lambda ^4 N_{*}}{H_{\rm inf}^2}}{6 f^2}\bigg)\cot \left(\frac{f \chi_{\rm osc}}{2}\right)\bigg)\bigg]+\rm constant.
\label{explicit chistar}
\end{equation}
In this calculation, the constant can be set to zero and the maxima of $\chi_{\rm osc}$ is around $0.08$ based on the periodic condition. It is worthwhile for plotting their relation after choosing suitable parameters in Planck units.
\begin{figure}[h!]
 \centering
  \includegraphics[]{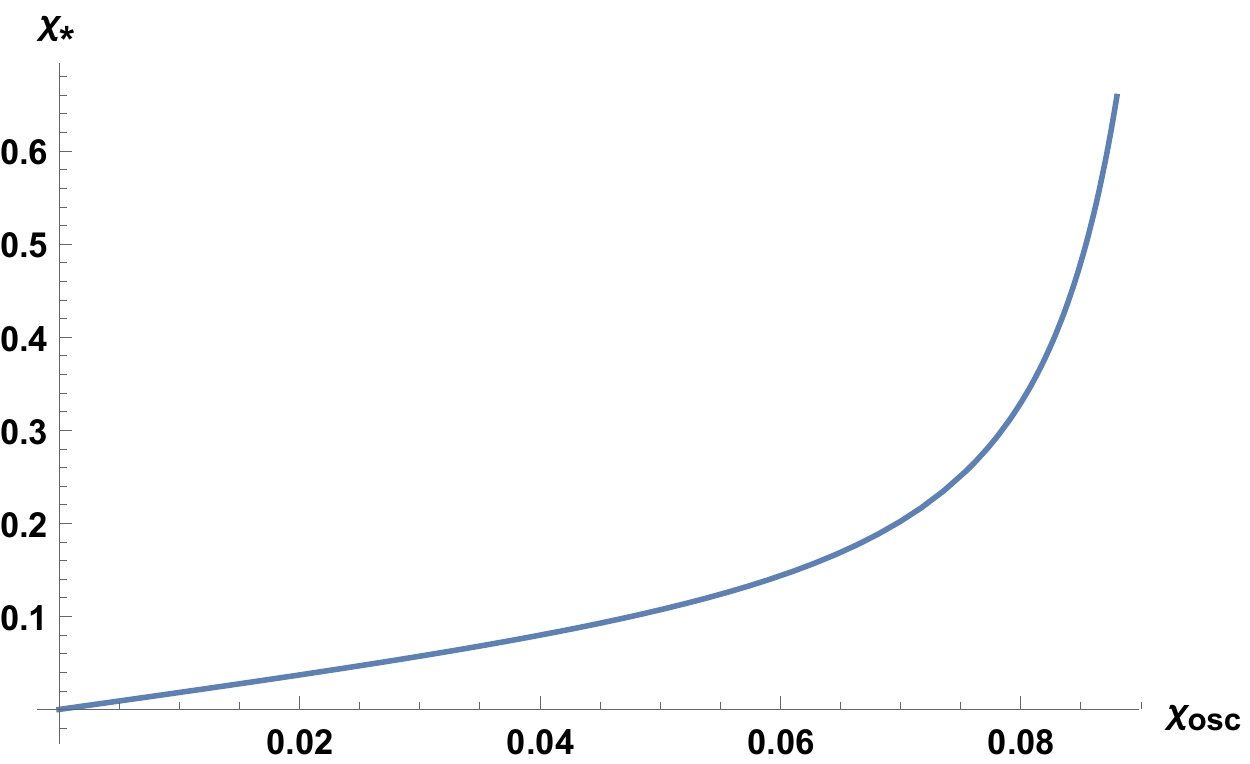}
\vskip -0.4cm
 \caption{ The relation of $\chi_{\rm oscc}$ and $\chi_*$ according to their explicit relation (\ref{explicit chistar}). During the whole range of $\chi_{\rm osc}$, its corresponding maximal value of $\chi_*$ is $0.7$. The parameters are setting as $N_*=50$, $f=3.36\times 10^{-2}$, $c=9/2$ (MD as an instance), $\Lambda=3.56\times 10^{-4}$ and $H_{\rm inf}=10^{-5}$ as adopting in Ref. \cite{Kawasaki:2011pd}.
 }
 \label{explicit formula for chistar and chiosc}
 \end{figure}
In figure \ref{explicit formula for chistar and chiosc}, it clearly indicates that the maximal value of curvaton field is approximating equalled to $0.7$ whose value is lighter than Planck mass at the horizon exit. Comparing with inflaton field, it is a light field making its energy density is subdominant during inflation. Once finding their explicit relation, we could find formula of $A$ corresponding to Pseudo-Nambu-Goldstone curvaton. Due to complication of formula of $A$, all of these formulas will be tackled by Mathematica. Being armed with these formulas, we will plot the non-linearity parameter in various epoches including second inflationary process, RD and MD. Being different with investigating chaotic curvaton, $A$ is also a function of $c$ whose various values corresponding to different periods. Due to this parameter, we cannot vary with $w$ to analyze nonlinearity parameter $f_{\rm NL}$. Finally we only study the individual case referring to specific $w$ and $c$.

Case $\mathbf{a}$: $w=-1$ and $c=3$

The explicit of $f_{\rm NL}$ is too complicated to express due to the complication of $A$. Actually, most curvaton models being with various potentials cannot find express $A$ explicitly since the relation between $\chi_{\rm osc}$ and $\chi_*$ is almost not possible, taking placing by the numerical methods as showing in Ref. \cite{Kawasaki:2011pd,Kobayashi:2012ba}. Once knowing these knowledge and meanwhile observing that $w=-1$ and $c=3$ will lead to the divergence of $f_{\rm NL}$ from Eq. (\ref{fnl for cur density2}). For better understanding this case, the plot will be given.
\begin{figure}[h!]
 \centering
  \includegraphics[height=8.8cm, width=8.52cm]{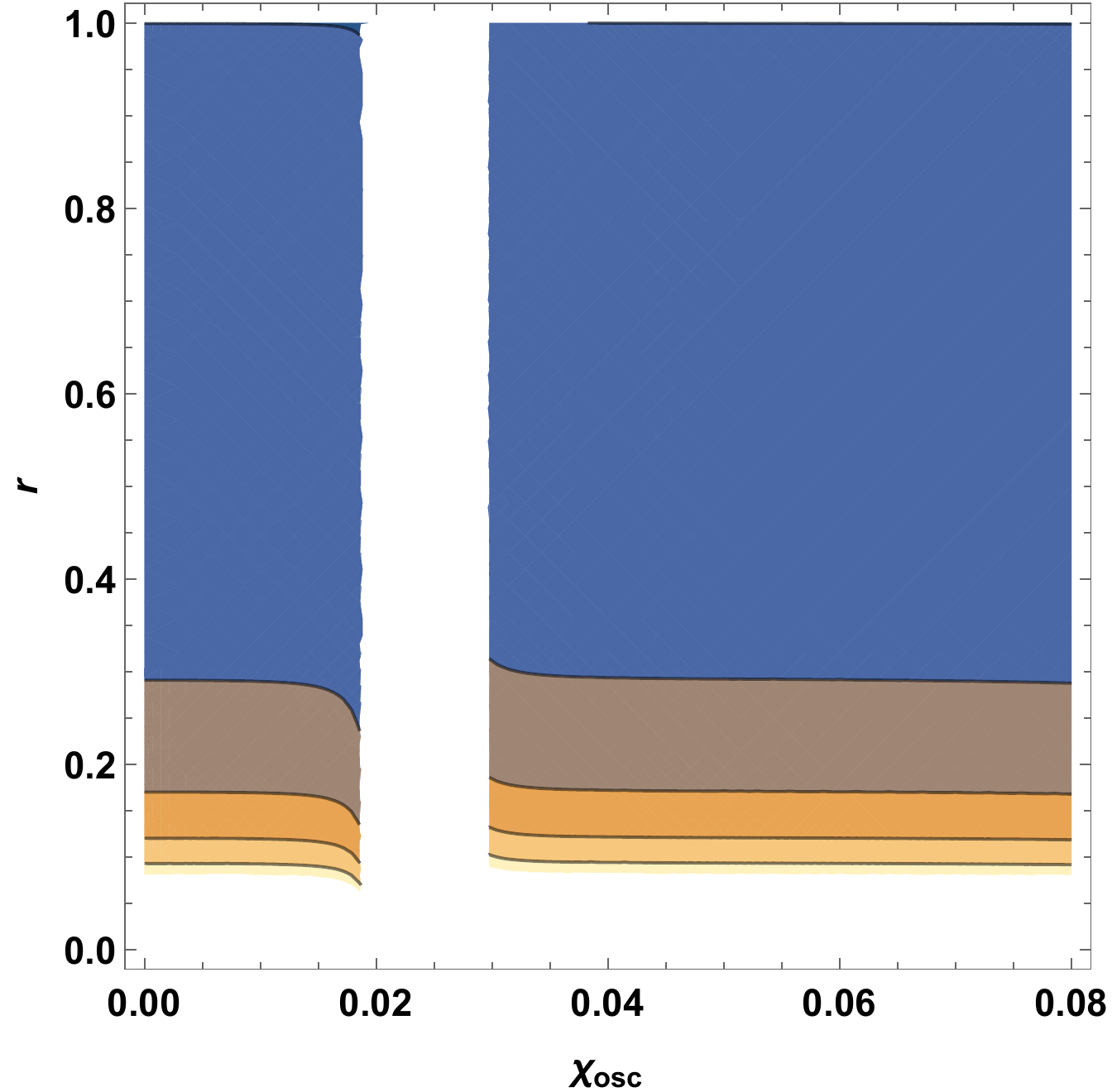}
\includegraphics[height=8.8cm, width=1.05cm ]{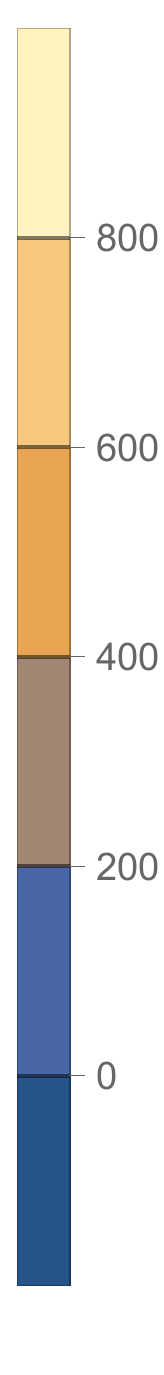}
\vskip -0.4cm
 \caption{ The horizontal line corresponds to $r_{\rm decay}$ whose range is $0 \leqslant r_{\rm decay}\leqslant 1$ including the whole possible value. The vertical line denotes the value of equation of $\chi_{\rm osc}$ locating from $0$ to $0.08$. The right panel shows that the value of $f_{\rm NL}$ matching its corresponding color. The parameters are set the same as figure \ref{explicit formula for chistar and chiosc}. }
 \label{fnl of axion for case a}
 \end{figure}
In figure \ref{fnl of axion for case a}, we could clearly see that the $f_{\rm NL}$ varying with $\chi_{\rm osc}$
and $r_{\rm decay}$. The observational constraint gives the upper limit whose value is less than $10$. From figure \ref{fnl of axion for case a}, it is almost impossible find this value, in particular, as $r_{\rm}<0.3$, $f_{\rm NL}$ already exceeds the upper limit of observational constraint. Additionally, there is also divergence as $\chi_{\rm osc}$ is between from $0.018$ to $0.03$. The varying trend of $f_{\rm NL}$ will flip as crossing these divergent areas. To sum up, the secondary inflation for curvaton will not happen in light of our discussion.

Case $\mathbf{a}$: $w=0$ and $c=9/2$

In this case, axionic curvaton behaves like pressureless matter. Its plot will also be gotten.
\begin{figure}[h!]
 \centering
  \includegraphics[height=8.8cm, width=8.52cm]{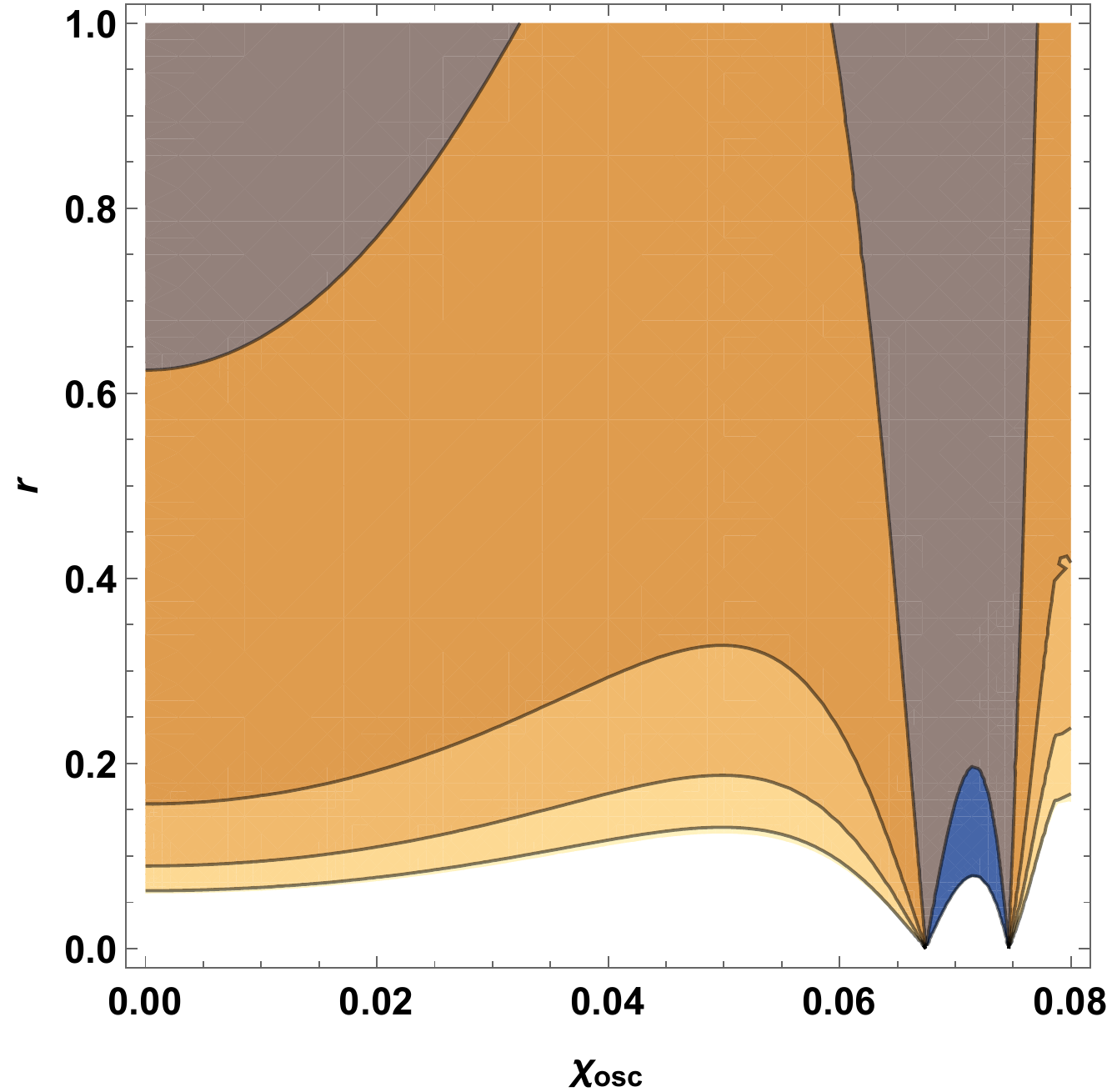}
\includegraphics[height=8.8cm, width=1.05cm ]{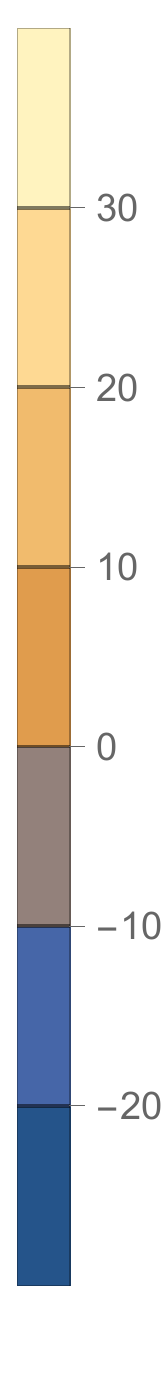}
\vskip -0.4cm
 \caption{ The horizontal line corresponds to $r_{\rm decay}$ whose range is $0 \leqslant r_{\rm decay}\leqslant 1$ including the whole possible value. The vertical line denotes the value of equation of $\chi_{\rm osc}$ locating from $0$ to $0.08$. The right panel shows that the value of $f_{\rm NL}$ matching its corresponding color. The parameters are set the same as figure \ref{explicit formula for chistar and chiosc}. }
 \label{fnl of axion for case b}
 \end{figure}
In figure \ref{fnl of axion for case b}, it clearly indicates that the most parameter space satisfied with observational constraints \cite{Ade:2015ava} especially for $r_{\rm decay}>0.2$. The value of $f_{\rm NL}$ will become negative as $0.067\leq \chi_{\rm osc}\leq 0.075$. If the observation could constrain the sign of $f_{\rm NL}$, it will give a strong constraints of our mechanism for curvaton. Comparing with Ref. \cite{Kawasaki:2011pd}, our formula is not not so highly depending on the field value of $\chi$, in which we use replace $\chi_*$ with $\chi_{\rm osc}$ to investigate. In this case, the upper limit $r_{\rm decay}$ is smaller comparing to chaotic curvaton, which means that fraction of curvaton among the total energy could be less even in MD.

case $\mathbf{c}$: $w=\frac{1}{3}, c=5$

In case, we will study the nonlinearity parameter in RD.
 \begin{figure}[h!]
 \centering
  \includegraphics[height=8.8cm, width=8.52cm]{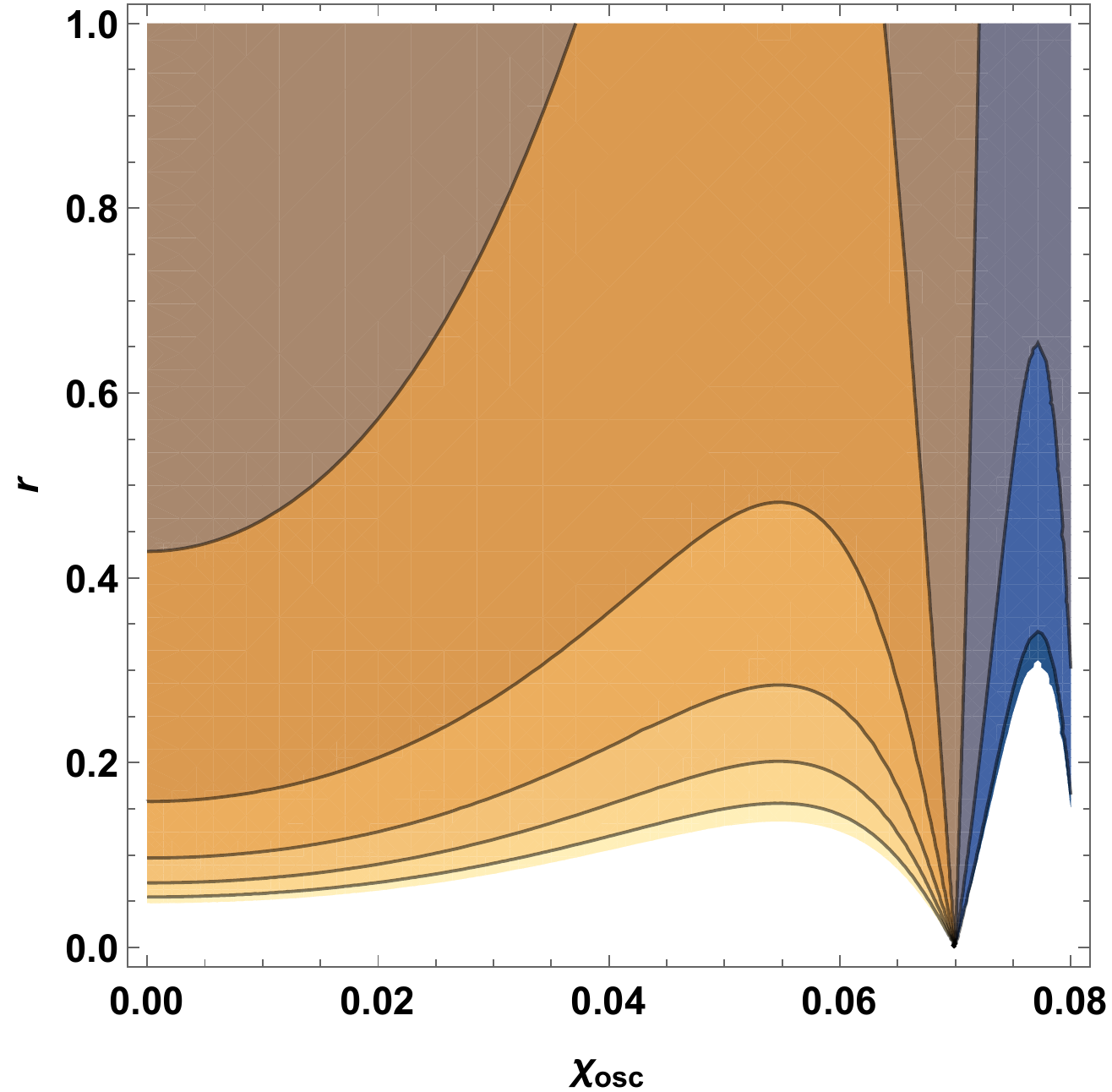}
\includegraphics[height=8.8cm, width=1.05cm ]{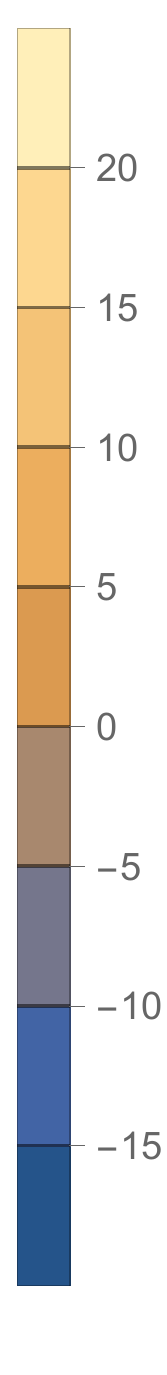}
\vskip -0.4cm
 \caption{ The horizontal line corresponds to $r_{\rm decay}$ whose range is $0 \leqslant r_{\rm decay}\leqslant 1$ including the whole possible value. The vertical line denotes the value of equation of $\chi_{\rm osc}$ locating from $0$ to $0.08$. The right panel shows that the value of $f_{\rm NL}$ matching its corresponding color. The parameters are set the same as figure \ref{explicit formula for chistar and chiosc}. }
 \label{fnl of axion for case c}
 \end{figure}
Generically, the trend of figure \ref{fnl of axion for case c}
is similar with figure \ref{fnl of axion for case b}. It contains lots of parameter spaces satisfied with observational constraints. The difference comes for the upper limit of $r_{\rm decay}$, its value is even smaller whose range could reach $0.1$, the discussion is the same since one could consider that curvaton is the production in MD as showing in our previous work \cite{Liu:2019xhn}. Another distinct place is that the sign of $f_{\rm NL}$ flips around $0.07\leq \chi_{\rm osc}\leq 0.08$.

In this section, we apply our extending $f_{\rm NL}$ to different curvaton models. Firstly, in light of framework \cite{Kawasaki:2011pd}, it has already known that power spectrum is not varying dramatically with energy scale. According to this point, we only concern the nonlinearity parameter $f_{\rm NL}$. Our findings are the generic curvaton mechanism that will not experience the second inflationary process, although there is tiny choice of parameter space for chaotic curvaton. As for a Pseudo-Nambu-Goldstone curvaton, our findings show that no matter what curvaton behaves as pressure or pressureless matter, most of parameter spaces satisfy with observational constraints \cite{Ade:2015ava}. The only differences are determined by their decay process, this point is illustrated in Ref. \cite{Kobayashi:2012ba} identified with comparison between $t_{\rm decay}$ and $t_{\rm reheating}$.

\section{Conclusion}
\label{conclusion}
In this paper, we have constructed a generalized $\delta N$ formalism consisting of merits of Ref. \cite{Cai:2010rt, Kawasaki:2011pd}. Our method could deal with curvaton models with generic potentials only requiring sinusoidal oscillation, meanwhile it can also handle curvaton mechanism in various periods explicitly showing by EoS $w$ (secondary inflation, MD, RD) with corresponding parameter $c$ in Section \ref{generalized delta n}. For achieving a successful curvaton mechanism in different era, Refs. \cite{Allahverdi:2006dr,Enqvist:2002rf} proposed the curvaton mechanism embedded into MSSM within the consideration of thermalization, in which the effect of thermalization could transfer the isocurvature perturbation into curvature perturbation ensuring the nearly scale invariant power spectrum. From another perspective. Ref. \cite{Kawasaki:2011pd} analyzes the non-Gaussianity associated with $f_{\rm NL}$. Although their method could work with different period (MD, RD, {\it e.t.c}), they simply assumed that the different epoch corresponds to the various values of $c$ by neglecting the contribution of kinetic term. It is unavoidable for wrongly estimating the precise contribution of kinetic terms. In order to compensate this flaw, we adopted the advantage of Ref. \cite{Cai:2010rt}, directly associated with EoS $w$, for investigation.

Once obtaining the key result for non-linearity parameter $f_{\rm NL}$ (\ref{fnl for cur density2}), we implement it into two curvaton models. One is the chaotic curvaton, the other one is the Pseudo-Nambu-Goldstone curvaton. In light of framework of \cite{Kobayashi:2012ba}, we only concern the non-Gaussianity identified with $f_{\rm NL}$ since the power power spectrum is nearly scale invariant in various models. In traditional curvaton scenario, it behaveas as pressless matter with $w=0$. However, we are not capable for distinguishing the period of occurrence of curvaton mechanism from the observations. Thus, curvaton mechanism can be fullfilled in various periods, $\it i.g.$ the curvaton mechanism is achieved by the decay of inflaton field \cite{Liu:2019xhn} during the preheating period, in which it may be realized in RD as inflaton practically decay. Furthermore, the variants of $\delta N$ formalism could proceed with various epoches of curvaton mechanism \cite{Cai:2010rt,Kawasaki:2011pd,Kobayashi:2012ba}. Accordingly, we discussed two specific curvaton models in distinct era. 

For the chaotic curvaton, we investigate the $f_{\rm NL}$. In the limit of $r_{\rm decay}\rightarrow 1$, $f_{\rm}\rightarrow -\frac{5}{4}$ nicely recovers the analysis of Ref. \cite{Sasaki:2006kq} in case $a$ of chaotic curvaton and $f_{\rm NL}$ will be divergent in the limit of $r_{\rm decay}\rightarrow 0$. For case $a$, it indicates that the secondary inflationary process is ruled out by observational constraints. However, the occurrence of second inflationary process will alive if there is a transition from DE era to MD showing in figure \ref{fnl with quadratic potential}.

The original curvaton mechanism assumed that it was an extra and independent field comparing to inflaton field. One possibility for accounting for its origin is Pseudo-Nambu-Goldstone curvaton. In this model, the value of $f_{\rm NL}$ shows the similar varying trend with chaotic curvaton as showing in figure \ref{fnl of axion for case c},~\ref{fnl of axion for case b} and \ref{fnl of axion for case a}. Due to the complication of $A$ written by Eq. (\ref{A}), we cannot transit $w$ from one era to another era taking place by parameter $c$. From these figures, it explicitly shows that most parameter spaces satisfy with observational constraints which determines the upper limit of $r_{\rm decay}>0.1$. And the case of $a$  will be ruled out by the observations.

Finally, we will emphasize the further validity of our new formula for $f_{\rm NL}$. For the traditional curvaton mechanism, curvaton corresponds to a pressureless matter with $w=0$, our formula will nicely recover the classical result $f_{\rm NL }=-5/4$ in the limit of $r_{\rm decay} \approx 1$. For its validity in RD, curvaton mechanism can be realized as inflaton decay. As our previous discussions mentioned, isocurvature perturbation could be transferred into curvature perturbation by considering the thermalization. This idea was proposed by \cite{Dvali:2003em,Enqvist:2003uk}, the inflaton coupling is not a constant anymore, subsequently it could translate into fluctuations in the reheating temperature. As a result, our formula of Eq. (\ref{fnl for cur density2}) is naturally applied into the framework of MSSM for curvaton model construction. Furthermore, we could also implement our method to explore the Non-Gaussianity in MSSM curvaton model. 

\section*{Acknowledgements}
LH is funded by Hunan Natural Science Foundation NO. 2020JJ5452 and Hunan Provincial Department of
Education, NO. 19B464. WL is funded by NSFC 1175012.

\bibliography{mybibfile}

\end{document}